\documentclass[%
 reprint,
 amsmath,amssymb,
 aps,
]{revtex4-2}

\usepackage{graphicx}
\usepackage{dcolumn}
\usepackage{bm}
\usepackage{float}
\usepackage{xcolor}

\newcommand{\sat}{\mathrm{sat}}

\begin{document}


\title{Effects of Finite Temperature and Pairing Correlations in Multi-$\Lambda$ Hypernuclei}

\author{Bahruz Suleymanli}
\affiliation{Physics Department, Yildiz Technical University, 34220 Esenler, Istanbul, Turkey}

\author{Kutsal Bozkurt}
\affiliation{Physics Department, Yildiz Technical University, 34220 Esenler, Istanbul, Turkey}
\affiliation{Universit\'e Paris-Saclay, CNRS/IN2P3, IJCLab, 91405 Orsay, France}

\author{Elias Khan}
\affiliation{Universit\'e Paris-Saclay, CNRS/IN2P3, IJCLab, 91405 Orsay, France}
\affiliation{Institut Universitaire de France (IUF)}

\author{Ha\c{s}im G\"{u}ven}
\affiliation{Universit\'e Paris-Saclay, CNRS/IN2P3, IJCLab, 91405 Orsay, France}
\affiliation{Physics Department, Yildiz Technical University, 34220 Esenler, Istanbul, Turkey}

\author{J\'er\^ome Margueron}
\affiliation{Univ Lyon, Univ Claude Bernard Lyon 1, CNRS/IN2P3, IP2I Lyon, UMR 5822, F-69622, Villeurbanne, France}
\affiliation{International Research Laboratory on Nuclear Physics and Astrophysics, Michigan State University and CNRS, East Lansing, MI 48824, USA} 

\date{\today}

\begin{abstract}
The influence of finite temperatures and pairing correlations on the ground state properties of multi-$\Lambda$ Ca, Sn and Pb hypernuclei is explored using finite temperature Hartree–Fock–Bogoliubov approach and contact pairing interaction. A critical temperature is predicted and is in agreement with the Bardeen-Cooper-Schrieffer relationship $k_B T_C^\Lambda \approx 0.5 \Delta^{T=0}_\Lambda$,
beyond which pairing correlations drop to zero.
Particle densities, $\Lambda$ single-particle energies, and nuclear radii are weakly impacted by pairing as well as by finite temperatures. However, other nuclear properties which are more sensitive to pairing correlations, such as $\Lambda$ pairing gaps, condensation energies, and abnormal densities are also more impacted by finite temperature, especially around the critical temperature. Furthermore, calculations show the occurrence of the pairing re-entrance effect in the $^{280}_{70\Lambda}$Pb hyperon-drip-line hypernucleus. Our study provides insight into the thermal evolution of $\Lambda$ pairing, i.e. the emergence and vanishing of  pairing correlations in multi-$\Lambda$ hypernuclei as a function of temperature.
\end{abstract}

\maketitle

\section{\label{sec:level1}Introduction}

Including the temperature effect in nuclear systems can have a large impact on astrophysical processes \cite{Ravlic:2023}. Hot nuclear matter exists only temporary in compact stars and in finite nuclei since thermal energy can be efficiently evacuated by neutrino and gamma-ray emissions. The properties of hot matter may however be different from the ones of cold systems such as neutron stars (NSs) or finite nuclei and may influence time evolution. For instance, proto-neutron stars (PNSs)  are transient celestial objects for which temperature impact dynamical evolution~\cite{Reddy:1999,Roberts:2012,Burrows:2013,Foglizzo:2015}.
During the first few seconds after their formation by massive star core-collapse (supernovae) or by binary neutron star mergers (kilonovae), matter can be heated to temperatures of the order of a few $10^{11}$~K, which drops down to about $T \lesssim 10^{10}$ K in less than an hour~\cite{Geppert2016}. During this cooling, hypernuclear matter may temporary exist in hot PNSs~\cite{1995AsA, YU201097c}, but the actual presence of hyperons in the inner core of PNSs crucially depends on the hyperon-hyperon interaction~\cite{SCHAFFNERBIELICH2010279}. Consequently, the interplay between the hyperon-hyperon interaction and temperature plays an important role in the understanding of the cooling properties of PNSs.

The uncertainties in the hyperon-hyperon interaction are however still too large to allow accurate predictions about hyperon matter. These uncertainties are largely due to the scarce amount of experimental data, since only a limited number of double-$\Lambda$ hypernuclei, including $^6_{\Lambda\Lambda}$He and $^{11}_{\Lambda\Lambda}$Be \cite{Collaboration:2013aa, PhysRevC.92.044313, 10.1093/ptep/pty149} have been produced and measured in laboratories. In the future, the production of hyperons at FAIR and JPARC will provide new data which will be employed to reduce the uncertainties in the actual modeling of hyperon matter \cite{Feliciello_2015, RevModPhys.88.035004, OHNISHI2020103773}.

Astrophysical data can also be employed to reduce the model uncertainties. For instance the presence of hyperons in NSs may impact their cooling, by influencing, neutrino emission processes. More precisely, two hyperon beta decays, namely $\Sigma^{-} \rightarrow \Lambda+e^{-}+\bar{\nu}$ and $\Lambda \rightarrow p+e^{-}+\bar{\nu}$, contributing to the direct URCA process, have been predicted to be very efficient in the long time ($\gg$years) cooling of NSs~\cite{10.1093/mnras/stab3126}. If these processes occur, the surface temperature would be lower than the observed ones~\cite{SCHAAB1996531,TSURUTA19981,10.1093/mnras/staa1871}. Therefore, it is necessary to reduce their efficiency, for instance, when the NS temperature is below the critical temperature $T_C^\Lambda$ for $\Lambda$ pairing~\cite{RevModPhys.64.1133, TSURUTA19981, YAKOVLEV20011,Schaab_1998}. If hyperons are present in NSs, the observed surface temperature is a sign for the existence of pairing among hyperons: in this superfluid state, the emissivity of strange decay reactions is reduced by a factor of exp$\left(-\Delta/kT\right)$, where $\Delta$ represents the pairing gap~\cite{TSURUTA19981}. 

In conclusion of these first remarks, the composition of dense matter in the core of NSs and PNSs has observational consequences which can be related to the properties of hyperon interactions. In addition, the impact of these interactions on the $\Lambda$ pairing gap is very important for the cooling of these stellar objects, and several theoretical predictions on $\Lambda$ pairing have been performed in baryonic matter
\cite{PhysRevC.57.409,Margueron:2018a,Lifshitz1980,TAKATSUKA2003C1003,doi:10.1142/S0218301314500785,PhysRevC.68.015801, 10.1143/PTP.115.355,PhysRevC.81.025801,Xu_2012}
and hypernuclei \cite{PhysRevC.94.054325,GULERIA201271, PhysRevC.94.024331, PhysRevC.107.044317,Samanta_2008, Samanta_2010,PhysRevC.95.024323, PhysRevC.97.034302, PhysRevC.95.034309,PhysRevC.98.014318, RONG2020135533, PhysRevC.105.034322}.

We briefly review these predictions. Based on a G-matrix, $\Lambda$ pairing gap has been predicted to be about a few $100$~keV~\cite{PhysRevC.57.409} for densities $n_\Lambda\approx n_\sat$, where $n_\sat$ is the saturation density of symmetric matter, $n_\sat\approx 0.155\pm0.005$~fm$^{-3}$~\cite{Margueron:2018a}, and the BCS relation obtained in the weak-coupling approximation $k_B T_C \approx 0.57 \Delta^{T=0}$ has been found to be well satisfied~\cite{Lifshitz1980,TAKATSUKA2003C1003,doi:10.1142/S0218301314500785}. Moreover, superfluidity has been predicted to occur at about $4n_\sat$ in beta-equilibrated matter, using one-boson-exchange potentials for the $\Lambda\Lambda$ interaction ~\cite{PhysRevC.68.015801, 10.1143/PTP.115.355}. Phenomenological nuclear interactions have also been employed to predict the properties of $\Lambda$ pairing in nuclear matter. For instance, a relativistic mean field (RMF) model has predicted the pairing gap to be 810~keV at the baryon density of 0.349 fm$^{-3}$~\cite{PhysRevC.81.025801}.
Other models consider $\sigma^*$ and $\varphi$ mesons (with strange quark) added on top of the other (without strange quark) mesons: the pairing gap is predicted to be reduced down to 282~keV at the total baryon density of 0.364 fm$^{-3}$~\cite{Xu_2012}.
These results illustrate the variability among different predictions.

One can also employ hypernuclei as a benchmark for $\Lambda\Lambda$ interactions. For instance, a study based on the relativistic Thomas-Fermi approach and applied to hypernuclei, has shown an interesting effect of the temperature: the $\Lambda$ radii increases faster as function of the temperature than the nucleon radii, since it is easier to excite the small amount of hyperons than the large number of nucleons~\cite{PhysRevC.94.054325}. Still in finite nuclei, Skyrme-Hartree-Fock model~\cite{GULERIA201271, PhysRevC.94.024331, PhysRevC.107.044317}, generalized liquid drop model \cite{Samanta_2008, Samanta_2010}, and beyond-mean-field SHF model \cite{PhysRevC.95.024323, PhysRevC.97.034302, PhysRevC.95.034309}, have been employed within the BCS approximation. Microscopic calculations applied to  multi-$\Lambda$ hypernuclei have emphasized the connection between pairing in $\Lambda$ channel and the $\Lambda \Lambda$ interaction \cite{PhysRevC.98.014318, RONG2020135533, PhysRevC.105.034322}. Using different methods to calibrate $\Lambda$ pairing strength, these studies have predicted the properties of multi-$\Lambda$ hypernuclei in their ground-state. However, there are no calculation in hypernuclei which explicitly treat temperature and $\Lambda$ pairing together. 

The main purpose of the present study is to understand the effect of temperature and $\Lambda$ pairing in hypernuclei, in view of the future production of a wealth of hypernuclei at FAIR and JPARC. It is also expected to impact hyperon matter in astrophysics. In this work, we extend our previous analysis of $\Lambda$ pairing in multistrange hypernuclei~\cite{PhysRevC.98.014318} by incorporating the finite temperature effect. To do so, we implement the finite temperature Hartree-Fock-Bogoliubov (FT-HFB) approach~\cite{GOODMAN198130}, where we consider Skyrme-type energy density functionals in the nucleon channel. To account for the $\Lambda$N interaction, we consider the density functionals DF-NSC89, DF-NSC97a, and DF-NSC97f, adjusted to reproduce Brueckner Hartree-Fock (BHF) results based on Nijmegen interactions NSC89, NSC97a and NSC97f~\cite{PhysRevC.62.064308, PhysRevC.64.044301}. To describe the $\Lambda \Lambda$ interaction, we use the empirical prescription EmpC~\cite{PhysRevC.96.054317}, which is adjusted to reproduce the binding energy of $^6_{\Lambda \Lambda}$He. 

In practice, we consider spherical symmetry in this first study, aiming at having a qualitative understanding of $\Lambda$ pairing at finite temperature. 
We have therefore specifically chosen hypernuclei with closed proton and neutron shells, such as ${ }_{-S \Lambda}^{40-\mathrm{S}} \mathrm{Ca},{ }_{-\mathrm{S} \Lambda}^{132-\mathrm{S}} \mathrm{Sn}$, and ${ }_{-\mathrm{S} \Lambda}^{208-\mathrm{S}} \mathrm{Pb}$, known for their semi-magic properties. Moreover, the structure of hyperon-drip-line hypernuclei is also examined, particularly those with open neutron shells located near closed neutron shells, such as $^{56,58,62,64}_{20\Lambda}$Ca, $^{168,170,174,176}_{40\Lambda}$Sn, and $^{274,276,280,282}_{70\Lambda}$Pb. Our study focuses on investigating the impact of the temperature on the low-energy properties of multi-$\Lambda$ hypernuclei. 

The present manuscript is organized as follows. In Sec.~\ref{sec:level2}, we provide a brief overview of the FT-HFB approach applied to multi-$\Lambda$ hypernuclei. In Sec.~\ref{sec:level3}, we explore the interplay between temperature and $\Lambda$ pairing, analyzing their impact on the low-energy properties. Conclusions are given in Sec.~\ref{sec:level4}.

\section{Finite-temperature Hartree–Fock–Bogoliubov Approach for Multi-$\Lambda$ Hypernuclei}
\label{sec:level2}

The present model is based on the FT-HFB approach, which is described in details in Ref.~\cite{GOODMAN198130}. It is a non-relativistic model for nucleons $\mathrm{N} = (p, n)$ and $\Lambda$ hyperons, which is employed to describe multi-$\Lambda$ hypernuclei. 

\subsection{Formalism}

\begin{table*} 
\tabcolsep=0.25cm
\def\arraystretch{1.4}
\caption{Parameters corresponding to the energy density and effective mass of the $\Lambda$ hyperons.} \label{tb:lambda_parameters}
\begin{tabular}{cccccccccccccc}
\hline \hline Functional & $\alpha_1$ & $\alpha_2$ & $\alpha_3$ & $\alpha_4$ & $\alpha_5$ & $\alpha_6$ & $\alpha_7$ & $\alpha_8$ & $\alpha_9$ & $\mu_1$ & $\mu_2$ & $\mu_3$ & $\mu_4$ \\
\hline DF-NSC89 + EmpC & 327 & 1159 & 1163 & 335 & 1102 & 1660 & 22.81 & 0 & 0 & 1 & 1.83 & 5.33 & 6.07\\
DF-NSC97a + EmpC & 423 & 1899 & 3795 & 577 & 4017 & 11061 & 21.12 & 0 & 0 & 0.98 & 1.72 & 3.18 & 0 \\
DF-NSC97f + EmpC & 384 & 1473 & 1933 & 635 & 1829 & 4100 & 33.25 & 0 & 0 & 0.93 & 2.19 & 3.89 & 0 \\
\hline \hline
\end{tabular}
\end{table*}

The total Hamiltonian reads
\begin{equation}
\widehat{H}=\sum_i \widehat{T}_i+\sum_{i,j} \widehat{H}^\prime_{ij},
\label{eq:hamiltonian}
\end{equation}
where $\widehat{T}$ represents the kinetic energy operators, $\widehat{H}^\prime$ denotes the interaction operator with $(i,j) = (\mathrm{N}, \mathrm{N})$, $(\mathrm{N}, \Lambda)$, and $(\Lambda, \Lambda)$.
The total energy 
can be expressed as 
\begin{equation}
E = \int \!\! d^3 r \sum_{i,j} \epsilon_{ij},
\label{eq:total_energy}
\end{equation}
with the following contributions: the hyperon-nucleon energy-density
\begin{eqnarray}
\epsilon_\mathrm{\Lambda N} &=& -(\alpha_1-\alpha_2 \rho_\mathrm{N}+\alpha_3 \rho_\mathrm{N}^2) \rho_{\Lambda} \rho_\mathrm{N}  \nonumber\\
&&+(\alpha_4-\alpha_5 \rho_\mathrm{N}+\alpha_6 \rho_\mathrm{N}^2) \rho_{\Lambda}^{5 / 3} \rho_\mathrm{N},
\label{eq:hyperon_nucleon}
\end{eqnarray}
the hyperon-hyperon energy-density
\begin{eqnarray}
\epsilon_\mathrm{\Lambda \Lambda} = \frac{\hbar^2}{2 m_\Lambda} \tau_\Lambda -(\alpha_7-\alpha_8 \rho_\Lambda+\alpha_9 \rho_\Lambda^2) \rho_{\Lambda}^2\, ,
\label{eq:hyperon_hyperon}
\end{eqnarray}
and the nucleon-nucleon energy-density, where $q$ stands for neutrons or protons,
\begin{eqnarray}
\epsilon_\mathrm{NN} &=& \frac{\hbar^2}{2 m_N} \tau_\mathrm{N} +  \frac{1}{2} t_0\left[\left(1+\frac{x_0}{2}\right) \rho_\mathrm{N}^2-\left(x_0+\frac{1}{2}\right) \sum_q \rho_q^2\right] \nonumber\\
&& +\frac{t_1}{4}\left\{\left(1+\frac{x_1}{2}\right)\left[\rho_\mathrm{N} \tau_\mathrm{N}+\frac{3}{4}(\nabla \rho_\mathrm{N})^2\right]-\left(x_1+\frac{1}{2}\right)\right.\nonumber\\
&& \left.\times \sum_q\left[\rho_q \tau_q+\frac{3}{4}\left(\nabla \rho_q\right)^2\right]\right\} \nonumber\\
&& +\frac{t_2}{4}\left\{\left(1+\frac{x_2}{2}\right) \left[\rho_\mathrm{N} \tau_\mathrm{N}-\frac{1}{4}(\nabla \rho_\mathrm{N})^2\right] +\left(x_2+\frac{1}{2}\right) \right.\nonumber\\
&& \left.\times \sum_q\left[\rho_q \tau_q-\frac{1}{4}\left(\nabla \rho_q\right)^2\right]\right\} \nonumber\\
&& -\frac{1}{16}\left(t_1 x_1+t_2 x_2\right) J_\mathrm{N}^2+\frac{1}{16}\left(t_1-t_2\right) \sum_q J_q^2 \nonumber\\
&& +\frac{1}{12} t_3 \rho_\mathrm{N}^\gamma\left[\left(1+\frac{x_3}{2}\right) \rho_\mathrm{N}^2-\left(x_3+\frac{1}{2}\right) \sum_q \rho_q^2\right] \nonumber\\
&& +\frac{1}{2} W_0\left(J_\mathrm{N} \nabla \rho_\mathrm{N}+\sum_q J_q \nabla \rho_q\right)+\epsilon_\mathrm{Coul}\, .
\label{eq:nucleon_nucleon}
\end{eqnarray}
The parameters $\alpha_{1-9}$ in Eqs.~\eqref{eq:hyperon_nucleon}  and \eqref{eq:hyperon_hyperon} are provided in Table \ref{tb:lambda_parameters} for the following functionals: DF-NSC89, DF-NSC97a, and DF-NSC97f \cite{PhysRevC.62.064308, PhysRevC.64.044301} with EmpC prescription for $\alpha_7$ \cite{PhysRevC.96.054317}. The density $\rho_\Lambda$ represents the $\Lambda$ density, and the nucleon density is $\rho_\mathrm{N} = \rho_p + \rho_n$. In infinite nuclear matter, the kinetic energy densities $\tau_\Lambda$ and $\tau_\mathrm{N}$ are analytical functions of the matter density: $\tau_i = \frac{3}{5} (6\pi^2/g_i)^{2/3} \rho_i^{5/3}$, where $g_i = 4 (2)$ for $i = \mathrm{N} (\Lambda)$. The parameters $t_{0-3}$, $x_{0-3}$, $\gamma$, and $W_0$ are given by the Skyrme interaction and $J$ represents the spin-current densities of nucleons. The term $\epsilon_\mathrm{Coul}$ corresponds to the Coulomb energy density. It is worth noting that in Eqs.~(\ref{eq:hyperon_nucleon}) and (\ref{eq:hyperon_hyperon}), we employ energy density functionals, which have been adjusted to reproduce Brueckner-Hartree-Fock predictions, incorporating both nucleons and $\Lambda$ hyperons \cite{PhysRevC.62.064308, PhysRevC.64.044301}, as well as the bond energy of $^6_{\Lambda \Lambda}$He. 

The HFB equations in coordinate representation are derived by taking the variation of the energy functional expressed in Eq.~(\ref{eq:total_energy}) with respect to the density matrices:
\begin{equation}
\left(\begin{array}{cc}
h - \lambda & \tilde{h} \\
\tilde{h} & -h + \lambda
\end{array}\right)\left(\begin{array}{l}
U_k \\
V_k
\end{array}\right) \\
=E_k\left(\begin{array}{c}
U_k \\
V_k
\end{array}\right)
\label{eq:hfb}
\end{equation}
where $h\left(\mathbf{r} \sigma, \mathbf{r}^{\prime} \sigma^{\prime}\right)=\frac{\delta E}{\delta \rho\left(\mathbf{r} \sigma, \mathbf{r}^{\prime} \sigma^{\prime}\right)}$ is the Hamiltonian containing the self-consistent field with particle density $\rho(r, r^\prime) = \frac{1}{4\pi} \sum_k (2j_k + 1) V_k (r) V_k^* (r^\prime)$ and the chemical potential $\lambda$, and $\tilde{h}\left(\mathbf{r} \sigma, \mathbf{r}^{\prime} \sigma^{\prime}\right)=\frac{\delta E}{\delta \tilde{\rho}\left(\mathbf{r} \sigma, \mathbf{r}^{\prime} \sigma^{\prime}\right)}$ is the pairing field with pairing density $\tilde{\rho} (r, r^\prime) = - \frac{1}{4\pi} \sum_k (2j_k + 1) V_k (r) U_k^* (r^\prime)$. The fields $U_k$ and $V_k$ are the components of the radial HFB wave function and $E_k$ is the quasiparticle energy. It should be noted that is that the considered model incorporates two distinct types of HFB mean fields potentials: 
\begin{equation}
V_{\Lambda}=\frac{\partial \left(\epsilon_{\Lambda \mathrm{N}} + \epsilon_{\Lambda \Lambda}\right)}{\partial \rho_{\Lambda}}-\left(\frac{m_{\Lambda}}{m_{\Lambda}^*\left(\rho_\mathrm{N}\right)}-1\right) \frac{\left(3 \pi^2\right)^{2 / 3}}{2 m_{\Lambda}} \rho_{\Lambda}^{2 / 3},
\label{eq:V_Lambda}
\end{equation}
and 
\begin{eqnarray}
V_N &=& \frac{\partial \left(\epsilon_{\Lambda \mathrm{N}} + \epsilon_\mathrm{NN}\right)}{\partial \rho_\mathrm{N}} +\frac{\partial}{\partial \rho_\mathrm{N}}\left(\frac{m_{\Lambda}}{m_{\Lambda}^*\left(\rho_\mathrm{N}\right)}\right)\nonumber\\
&& \times \left(\frac{\tau_{\Lambda}}{2 m_{\Lambda}}-\frac{3}{5} \frac{\left(3 \pi^2\right)^{2 / 3}}{2 m_{\Lambda}} \rho_{\Lambda}^{5 / 3}\right)\, .
\label{eq:V_N}
\end{eqnarray}
The $\Lambda$ effective mass reads
\begin{equation}
\frac{m_{\Lambda}^*\left(\rho_\mathrm{N}\right)}{m_{\Lambda}}=\mu_1-\mu_2 \rho_\mathrm{N}+\mu_3 \rho_\mathrm{N}^2-\mu_4 \rho_\mathrm{N}^3,
\label{eq:effective_mass}
\end{equation}
where the parameters $\mu_{1-4}$ are given in Table~\ref{tb:lambda_parameters}.
The $\Lambda \Lambda$ pairing interaction is defined as a volume-type contact interaction, similar to the one used in Ref.~\cite{PhysRevC.98.014318}:
\begin{equation}
\label{pair_potential}
V_{\Lambda_\text{pair}} = V_{\Lambda_0} \delta(\bm{r}_1 - \bm{r}_2)\, ,
\end{equation}
where $V_{\Lambda_0}$ is the $\Lambda \Lambda$ pairing interaction strength.

All equations presented so far, including Eq.~(\ref{eq:hfb}), are identical at finite temperature~\cite{GOODMAN198130, PhysRevC.70.025801} provided the densities are generalised at finite temperature according to the following expressions:
\begin{eqnarray}
\rho_T(r) &=& \frac{1}{4 \pi} \sum_k\left(2 j_k+1\right)\nonumber\\
&& \times \left[V_k^*(r) V_k(r)\left(1-f_k\right)+U_k^*(r) U_k(r) f_k\right]\, , \label{eq:rho_T} \\
\tau_T(r) &=&  \frac{1}{4 \pi} \sum_k\left(2 j_k+1\right)\nonumber\\
&& \times \left\{\left[\left(\frac{d V_k}{d r}-\frac{V_k}{r}\right)^2+\frac{l_k\left(l_k+1\right)}{r^2} V_k^2\right]\left(1-f_k\right)\right. \nonumber\\
&&   +  {\left.\left[\left(\frac{d U_k}{d r}-\frac{U_k}{r}\right)^2+\frac{l_k\left(l_k+1\right)}{r^2} U_k^2\right] f_k\right\} }\, ,
\label{eq:tau_T} \\
J_T(r) &=&  \frac{1}{4 \pi} \sum_k\left(2 j_k+1\right)\left[j_k\left(j_k+1\right)-l_k\left(l_k+1\right)-\frac{3}{4}\right] \nonumber\\
&&\left\{V_k^2\left(1-f_k\right)+U_k^2 f_k\right\},
\label{eq:J_T}
\end{eqnarray}
where $f_k = \left[1+\exp\left(E_k/k_B T\right)\right]^{-1}$ is the Fermi distribution, $k_B$ is the Boltzmann constant, and $T$ is the temperature. The low-energy properties in the equilibrium state at finite temperature for multi-$\Lambda$ hypernuclei are obtained from the solution of Eq.~(\ref{eq:hfb}), injecting Eqs.~(\ref{eq:rho_T})-(\ref{eq:J_T}) into Eqs.~(\ref{eq:total_energy})-(\ref{eq:effective_mass}).

\subsection{Numerical Strategy}

\begin{table}
\tabcolsep=0.1cm
\def\arraystretch{1.4}
\caption{$\Lambda \Lambda$ pairing interaction strength and corresponding hyperisotopic chain for each functional.} \label{tb:strength}
\begin{tabular}{ccc}
\hline \hline Functional & $\begin{array}{c}\text { V$_{\Lambda_0}$} \\
\left(\mathrm{MeV} \cdot \mathrm{fm}^3\right)\end{array}$ & $\begin{array}{c} \text{Hyperisotopic chain}\end{array}$ \\
\hline DF-NSC89+EmpC & -139 & $^{40-S}_{-S\Lambda}$Ca $(-S = 6 - 20)$ \\
DF-NSC97a+EmpC & -148 &  $^{40-S}_{-S\Lambda}$Ca $(-S = 6 - 20)$ \\
DF-NSC97f+EmpC & -180 & $^{40-S}_{-S\Lambda}$Ca $(-S = 6 - 20)$ \\
DF-NSC89+EmpC & -158 &  $^{132-S}_{-S\Lambda}$Sn $(-S = 18 - 40)$ \\
DF-NSC97a+EmpC & -145 & $^{132-S}_{-S\Lambda}$Sn $(-S = 18 - 40)$\\
DF-NSC97f+EmpC & -180 &  $^{132-S}_{-S\Lambda}$Sn $(-S = 18 - 40)$ \\
DF-NSC89+EmpC & -184 & $^{208-S}_{-S\Lambda}$Pb $(-S = 58 - 70)$  \\
DF-NSC97a+EmpC & -180 & $^{208-S}_{-S\Lambda}$Pb $(-S = 58 - 70)$ \\
DF-NSC97f+EmpC & -220 & $^{208-S}_{-S\Lambda}$Pb $(-S = 58 - 70)$ \\
\hline \hline
\end{tabular}
\end{table}

The FT-HFB equations for multi-$\Lambda$ hypernuclei presented hereabove are solved in coordinate representation and considering spherical symmetry. The SLy5 Skyrme interaction is fixed for the nucleon sector since it accurately reproduces stable and exotic nuclei properties~\cite{CHABANAT1998231}.
As usual, the spin-orbit term in the $\Lambda$ mean field channel is neglected, see Ref.~\cite{FINELLI200790} for instance. 

Following the approach suggested in Ref.~\cite{PhysRevC.98.014318}, the strength of the $\Lambda\Lambda$ pairing interaction $V_{\Lambda_0}$ is adjusted to reproduce the maximum value of the theoretical BCS prediction in uniform matter calculated in Ref.~\cite{PhysRevC.68.015801}, see Tab.~\ref{tb:strength}. It should be noted that the pairing gap at T=0 in nuclei are expected to be close to the one of uniform matter. For instance, in the considered $-S$ intervals displayed in Tab.~\ref{tb:strength}, the average pairing gap $\bar{\Delta}_\Lambda$ (running over a set of hypernuclei changing the number of hyperons while keeping the number of protons and neutrons) is close to the value in uniform matter~\cite{PhysRevC.98.014318}.

We employ the Numerov method to solve the FT-HFB equations and we consider the Dirichlet boundary condition for vanishing wave functions. The wave-functions are obtained numerically for a spherical box with a radius of 30~fm and a resolution of 0.1~fm. The self-consistent solution is obtained by an iterative process, which is repeated until the total energy converges within an accuracy of about $10^{-8}$~MeV.

\section{Results at finite temperatures}
\label{sec:level3}

In this section, we investigate how the pairing gaps, condensation energies, normal and pairing densities, and radii of multi-$\Lambda$ hypernuclei evolve as function of the temperature, considering different numbers of $\Lambda$ hyperons $(-S)$ present in the system. 
We focus on studying the hyper-isotopes $^{40-S}_{-S\Lambda}$Ca, $^{132-S}_{-S\Lambda}$Sn, and $^{208-S}_{-S\Lambda}$Pb, being magic in the nucleon sector.  Additionally, in cases where the hypernuclei are close to the hyperon-drip-line, investigations have also explored hypernuclei with neutron open-shell configurations at neutron numbers lying between $\pm 2$ and $\pm 4$ from the neutron magic number.

\subsection{$\Lambda$ Pairing Gaps and Binding Energies}

\begin{figure}
\includegraphics[width=0.4681\textwidth]{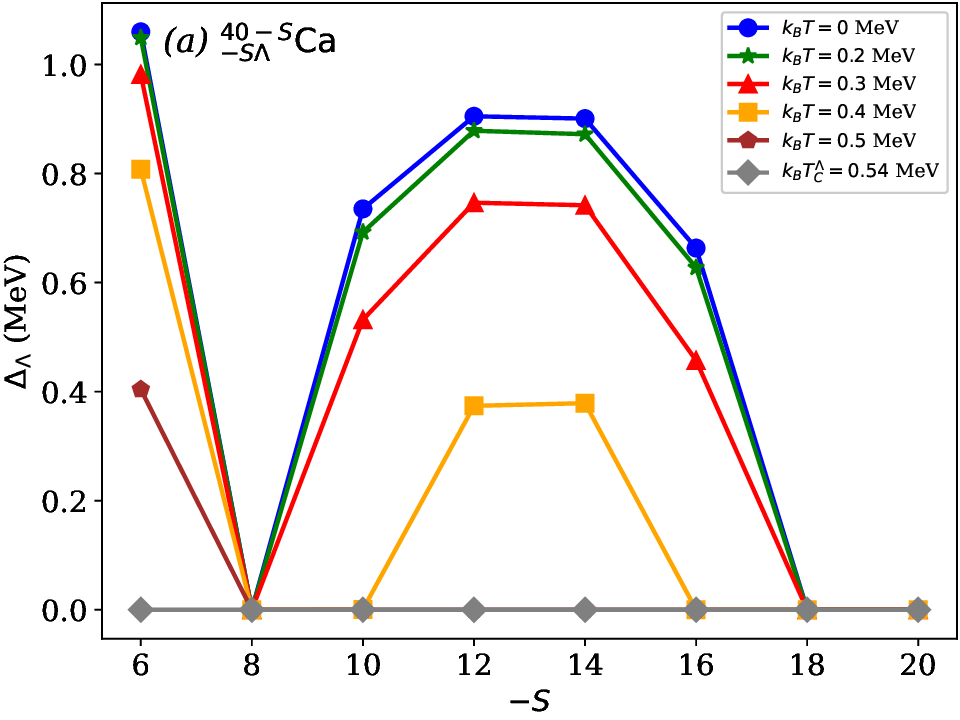}
\hspace{\fill}
\includegraphics[width=0.4681\textwidth]{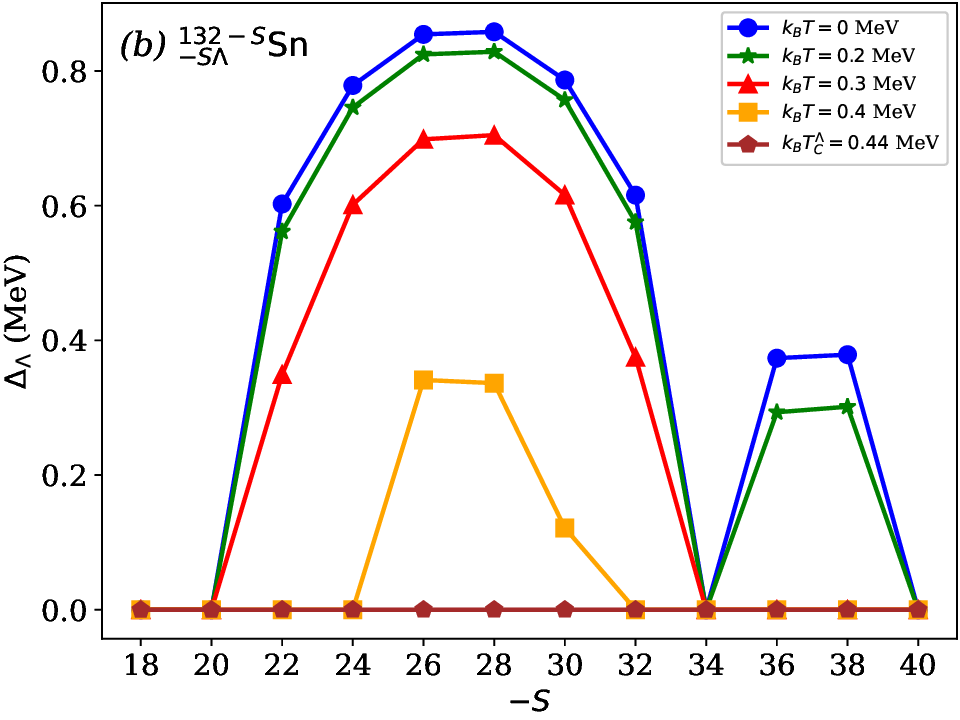}
\hspace{\fill}
\includegraphics[width=0.4681\textwidth]{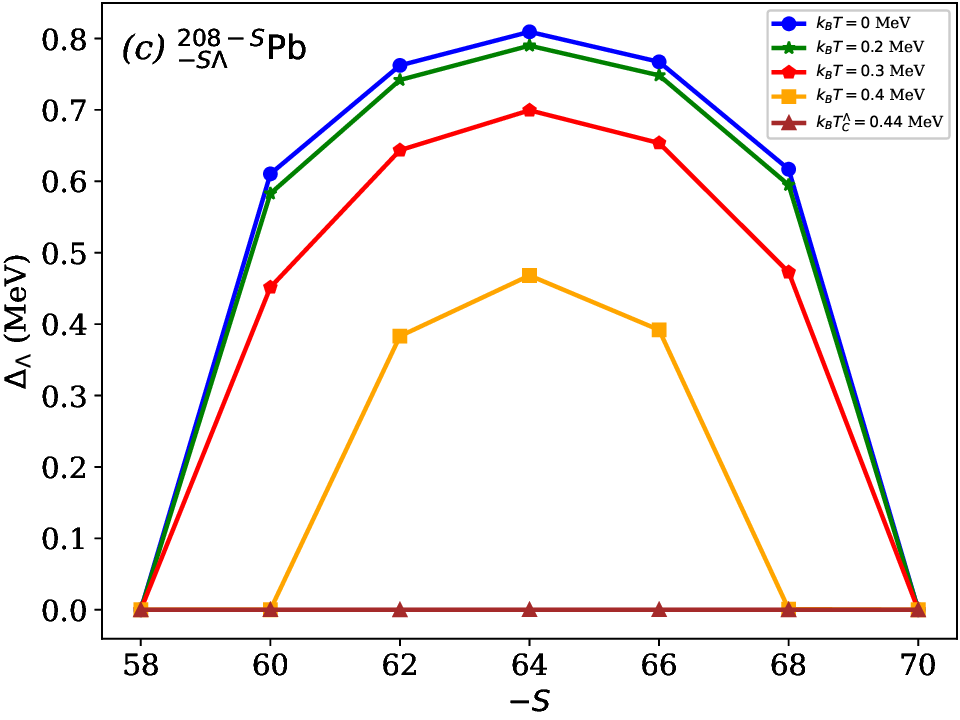}
\caption{Evolution of $\Lambda$ pairing gap with increasing temperature for $^{40-S}_{-S\Lambda}$Ca (a), $^{132-S}_{-S\Lambda}$Sn (b), and $^{208-S}_{-S\Lambda}$Pb (c) hypernuclei using DF-NSC97a+EmpC functional.}
\label{fig:gap_97ae}
\end{figure}

We first study the $\Lambda$ pairing gap ($\Delta_\Lambda$) as function of the temperature in the following multi-$\Lambda$ hyper-isotopes: $^{40-S}_{-S\Lambda}$Ca, $^{132-S}_{-S\Lambda}$Sn, and $^{208-S}_{-S\Lambda}$Pb. The results for $\Delta_\Lambda(T)$ are shown
in Fig.~\ref{fig:gap_97ae}, using the DF-NSC97a+EmpC functional. 

\begin{figure}
\includegraphics[width=0.4681\textwidth]{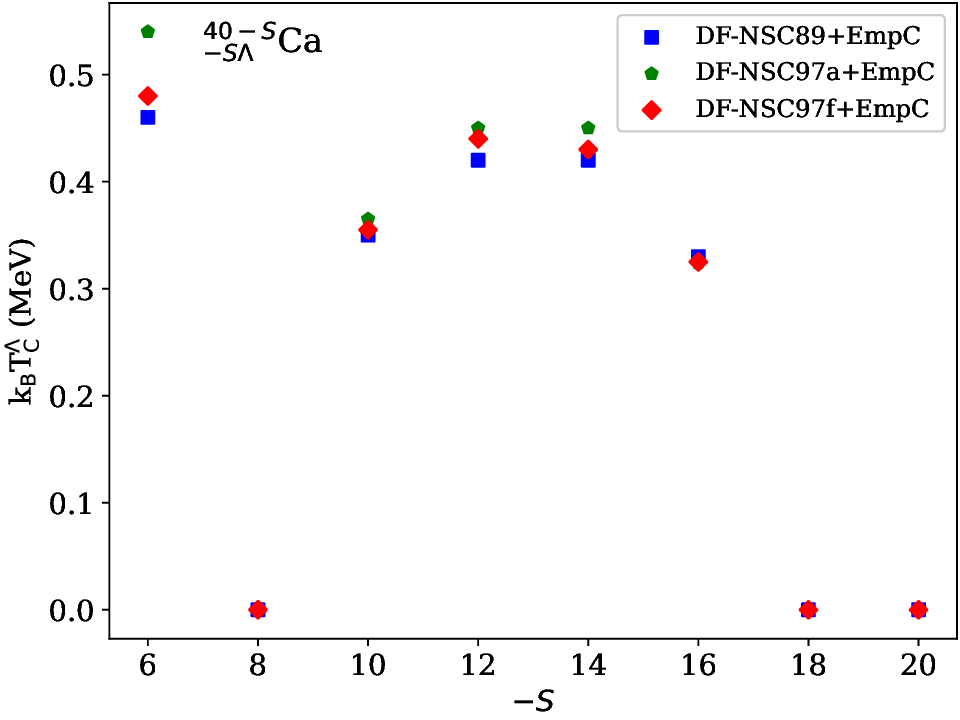}
\caption{Impact of strangeness on finite temperature for $^{40-S}_{-S\Lambda}$Ca hypernuclei.}
\label{fig:t_c}
\end{figure}

As the temperature increases, the pairing gaps $\Delta_\Lambda$ decrease until temperature reaches the critical temperature $T_C^\Lambda$ for $\Lambda$ pairing, which is studied in more details in the following.
At low temperature, the pairing gap is only weakly reduced but it drops down only when the temperature is close to the $T_C^\Lambda$. Note that for hyper-nuclei with $-S$ close to magic numbers, the pairing gap drops to zero even before the critical temperature is reached, at about $k_BT=0.4$~MeV. This trend is visible in Fig.~\ref{fig:t_c}, illustrating the influence of strangeness on finite temperature across various functionals for $^{40-S}_{-S\Lambda}$Ca hypernuclei.
In the three panels of Fig.~\ref{fig:gap_97ae} the largest temperature is set to the maximal critical temperature $T_C^\Lambda$ for the considered hyperisotopic chain (e.g. 0.54 MeV in the $^{40-S}_{-S\Lambda}$Ca case). 

In $^{40-S}_{-S\Lambda}$Ca, the $\Lambda$ pairing gap $\Delta_\Lambda$ is maximum at $-S = 6$ in the studied range $-S = 6-20$. Similarly, for $^{132-S}_{-S\Lambda}$Sn, the maximum pairing gap is obtained at $- S = 28$ in the range $-S = 18-40$, and for $^{208-S}_{-S\Lambda}$Pb, it occurs at $-S = 64$ in the range $- S = 58-70$. 
The pairing gaps are quenched for the following magic numbers: $-S = 8$, $18$, $20$, $34$, $40$, $58$, and $70$. This sequence is different from the usual one in non-strange nuclei since the spin-orbit has a negligible impact in the $\Lambda$ channel, involving a different shell structure. 

\begin{figure}
\includegraphics[width=0.4681\textwidth]{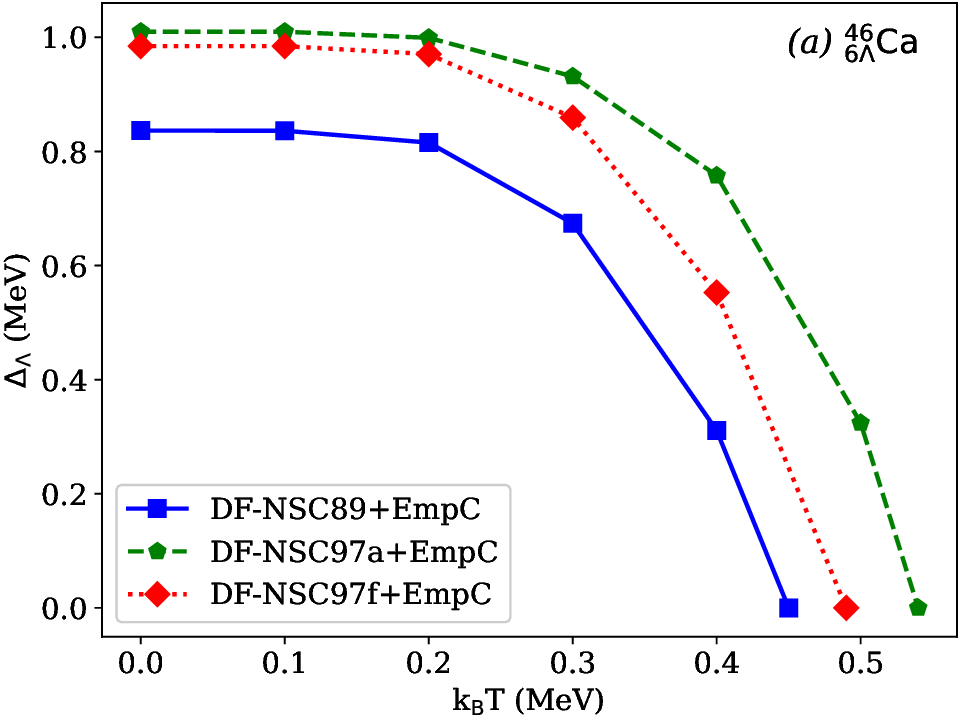}
\hspace{\fill}
\includegraphics[width=0.4681\textwidth]{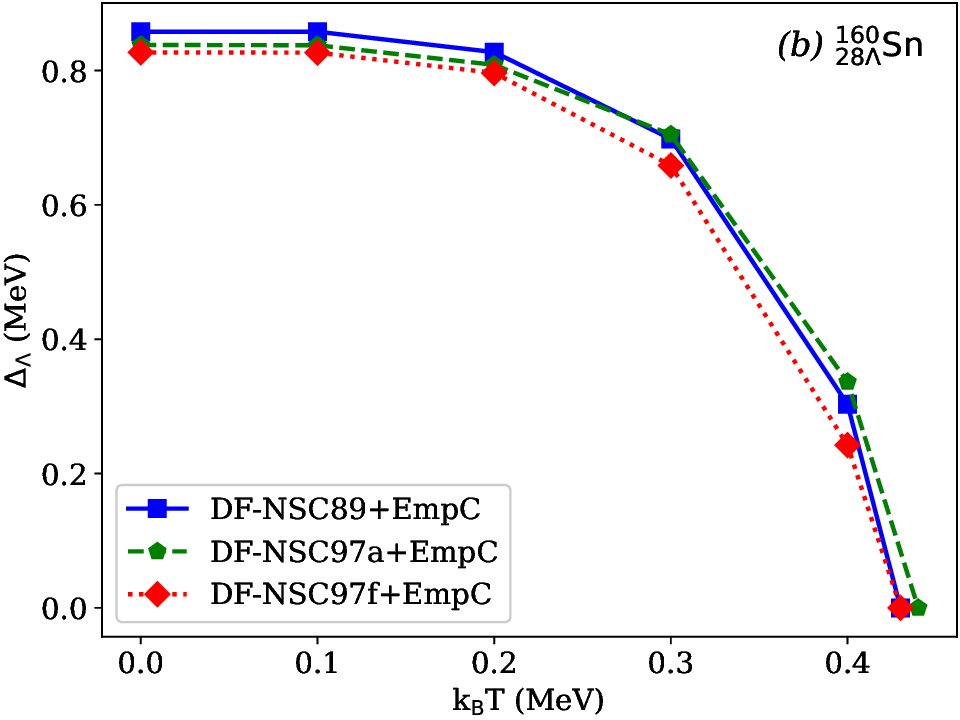}
\hspace{\fill}
\includegraphics[width=0.4681\textwidth]{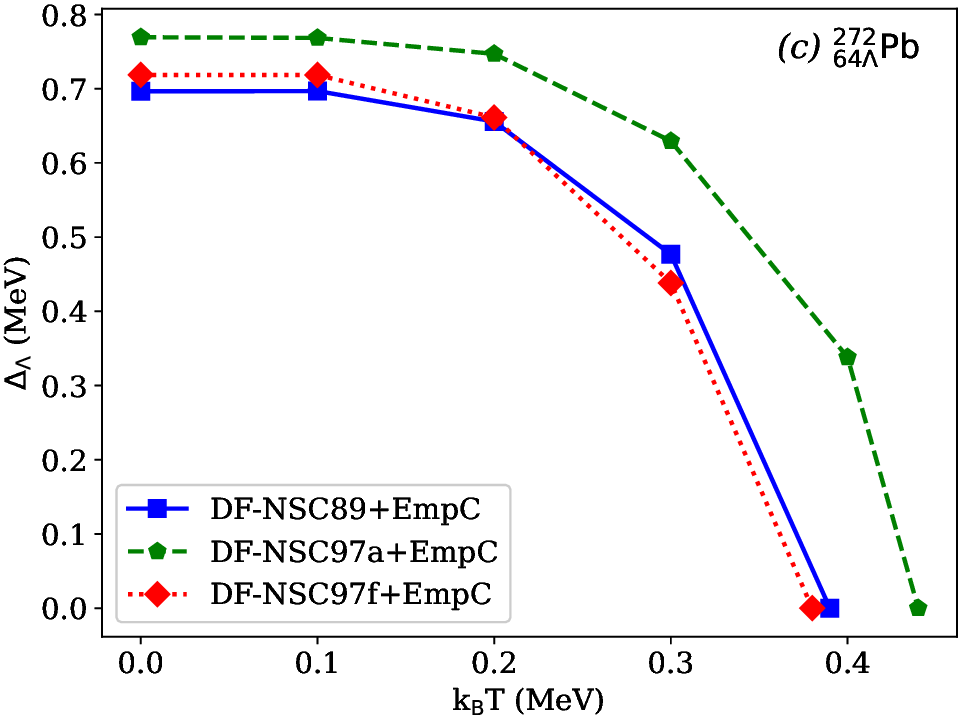}
\caption{Evolution of $\Lambda$ pairing gap with increasing temperature for $^{46}_{6\Lambda}$Ca (a), $^{160}_{28\Lambda}$Sn (b), and $^{272}_{64\Lambda}$Pb (c) hypernuclei. The plot depicts the results obtained from three functionals: DF-NSC89+EmpC (solid line), DF-NSC97a+EmpC (dashed line), and DF-NSC97f+EmpC (dotted line).}
\label{fig:p_gaps}
\end{figure}

It should be noted that qualitatively similar results are obtained for DF-NSC89+EmpC and DF-NSC97f+EmpC functionals. This can be seen in Fig.~\ref{fig:p_gaps} for $^{46}_{6\Lambda}$Ca, $^{160}_{28\Lambda}$Sn, and $^{272}_{64\Lambda}$Pb hypernuclei. Comparing $\Delta_{\Lambda}$ values for all three functionals, the DF-NSC97a+EmpC functional exhibits the largest $\Lambda$ pairing gap value across almost all temperatures. While a larger density of states around the Fermi level generally contributes to enhanced pairing gap, for the hypernuclei we are looking at, the increased number of bound single-particle states expands the available states for pairing interactions. This leads to a larger pairing gap, driven by the deeper $N \Lambda$ potential of the DF-NSC97a+EmpC functional compared to DF-NSC89+EmpC and DF-NSC97f+EmpC \cite{Collaboration:2013aa, PhysRevC.92.044313}.

Table~\ref{tb:critical} displays the values of the pairing gaps in the ground-state 
$\Delta_\Lambda^{T=0}$ and the critical temperature $T_C^\Lambda$ for a set of hypernuclei. We find that the BCS relation in uniform matter, namely $k_B T_C \approx 0.5 \Delta_\Lambda^{T=0}$, is well reproduced for finite hypernuclei. Note that the BCS relation is almost independent of the functional, since all used functionals impact both $\Delta_\Lambda^{T=0}$ and $T_C^\Lambda$.

\begin{table}
\tabcolsep=0.2cm
\def\arraystretch{1.4}
\caption{Pairing gap in the ground-state $\Delta_\Lambda^{T=0}$ and critical temperatures $T_C^\Lambda$ for different functionals and for a set of hypernuclei, which are the ones with maximum pairing gap, see Fig.~\ref{fig:gap_97ae}.} \label{tb:critical}
\begin{tabular}{cccc}
\hline \hline Functional & Hypernucleus & $\begin{array}{c}\text { $\Delta_\Lambda^{T=0}$} \\
\left(\mathrm{MeV}\right)\end{array}$ & $\begin{array}{c}\text { k$_\mathrm{B}$T$_\mathrm{C}^\Lambda$} \\
\left(\mathrm{MeV}\right)\end{array}$  \\
\hline DF-NSC89+EmpC & $^{46}_{6\Lambda}$Ca & 0.82 & 0.46  \\
DF-NSC97a+EmpC & $^{46}_{6\Lambda}$Ca & 1.04 & 0.54  \\
DF-NSC97f+EmpC & $^{46}_{6\Lambda}$Ca & 0.98 & 0.48 \\
DF-NSC89+EmpC & $^{160}_{28\Lambda}$Sn & 0.84 &  0.43 \\
DF-NSC97a+EmpC & $^{160}_{28\Lambda}$Sn & 0.82 & 0.44 \\
DF-NSC97f+EmpC & $^{160}_{28\Lambda}$Sn & 0.82 & 0.43\\
DF-NSC89+EmpC & $^{272}_{64\Lambda}$Pb & 0.69 &  0.39 \\
DF-NSC97a+EmpC & $^{272}_{64\Lambda}$Pb & 0.76 & 0.44  \\
DF-NSC97f+EmpC & $^{272}_{64\Lambda}$Pb & 0.71 &  0.38 \\
\hline \hline
\end{tabular}
\end{table}

\begin{figure}
\centering
\includegraphics[width=0.4681\textwidth]{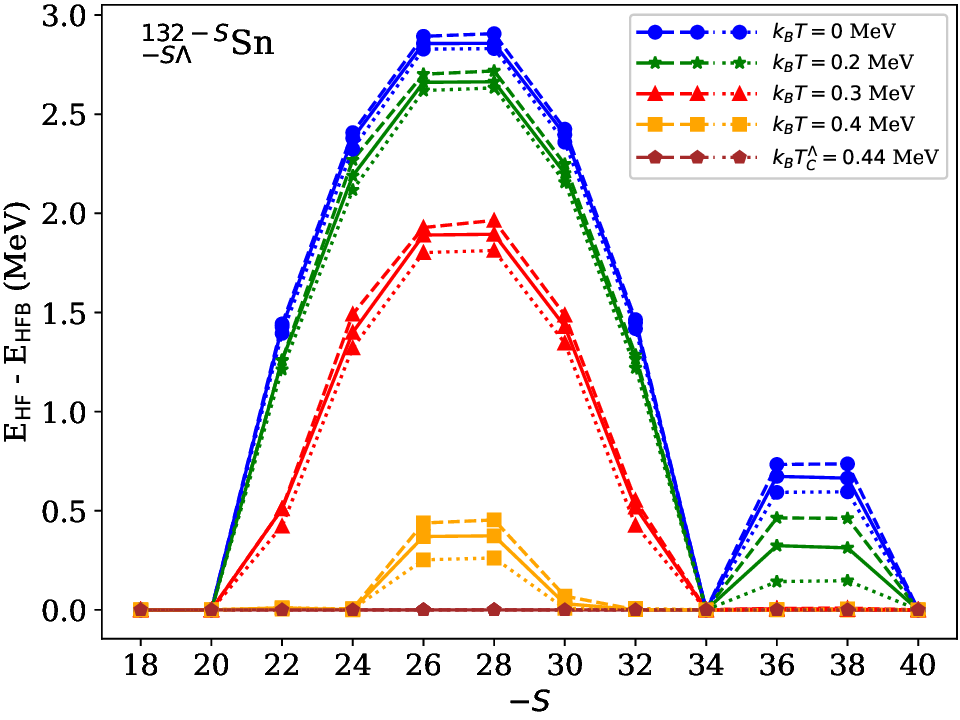}
\caption{Temperature-dependent binding energy differences of HF and HFB approaches for $^{132-S}_{-S\Lambda}$Sn hypernuclei with different functionals. The plot depicts the results obtained from three functionals: DF-NSC89+EmpC (solid line), DF-NSC97a+EmpC (dashed line), and DF-NSC97f+EmpC (dotted line).}\label{fig:binding_energy}
\end{figure}

The effects of finite temperatures and pairing correlations on binding energy can also be assessed by estimating the condensation energy, defined as E$_\text{cond}$ = E$_\text{HF}$ - E$_\text{HFB}$. Fig.~\ref{fig:binding_energy} shows E$_\text{cond}$ for a set of $^{132-S}_{-S\Lambda}$Sn hypernuclei, using DF-NSC89+EmpC, DF-NSC97a+EmpC and DF-NSC97f+EmpC functionals. The behavior of the condensation energy closely mirrors the trend observed in the $\Lambda$ pairing gap of Fig.~\ref{fig:gap_97ae}.

The peak value of the condensation energy nearly reaches $3$ MeV at $T = 0$. With increasing temperature, the E$_\text{HF}$ - E$_\text{HFB}$ value experiences a consistent reduction across all hypernuclei, mirroring the pattern displayed in Fig.~\ref{fig:gap_97ae}. For instance, as the temperature increases to $k_B T = 0.3$ MeV, the peak of the condensation energy decresases to approximately 2 MeV. At a specific temperature threshold E$_\text{HF}$ = E$_\text{HFB}$ across the whole $-S$ range.

\subsection{Normal and abnormal pairing densities}

\begin{figure}
\includegraphics[width=0.4681\textwidth]{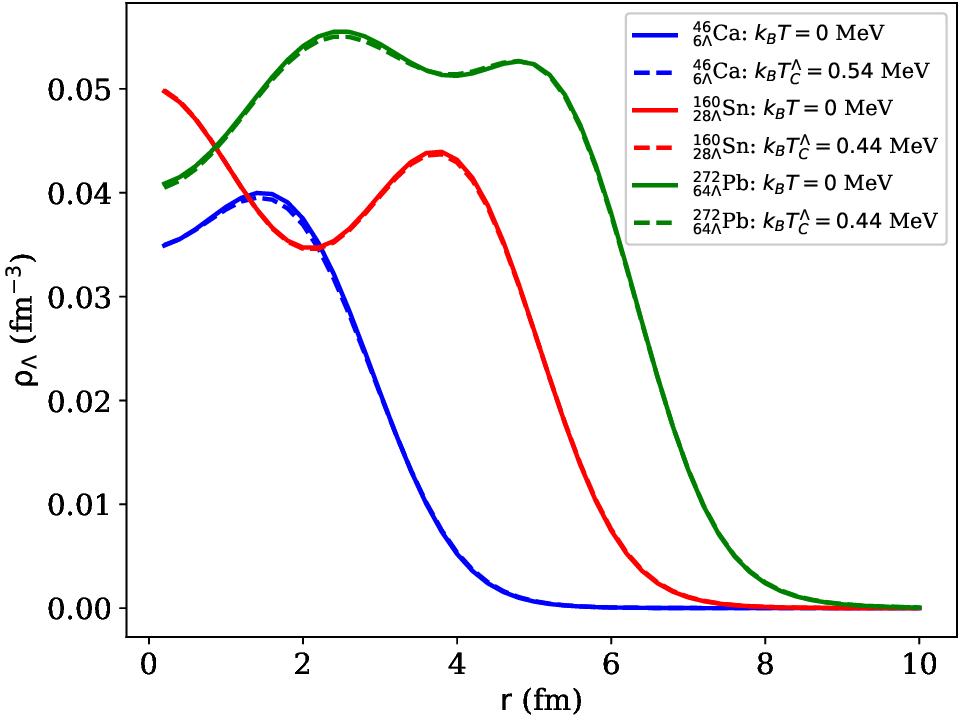}
\caption{Normal density profiles of $^{46}_{6\Lambda}$Ca, $^{160}_{28\Lambda}$Sn, and $^{272}_{64\Lambda}$Pb hypernuclei at finite temperatures using DF-NSC97a+EmpC functional.}\label{fig:normal_densities}
\end{figure}

Fig.~\ref{fig:normal_densities} displays the density profiles for typical three hyper-nuclei, $^{46}_{6\Lambda}$Ca, $^{160}_{28\Lambda}$Sn, and $^{272}_{64\Lambda}$Pb, for $T=0$ and $T=T_C^\Lambda$.
The impact of temperature on the normal densities is very small in the range of temperature bounded by the critical temperature. The $\Lambda$ single-particle spectrum has also been analyzed.
In the considered hypernuclei, shifts in energy levels of up to a few hundred keV are observed, as temperature increases. 
Therefore, there isn't an important impact of the temperature of the $\Lambda$ single-particle spectrum.

\begin{figure}
\centering
\includegraphics[width=0.4681\textwidth]{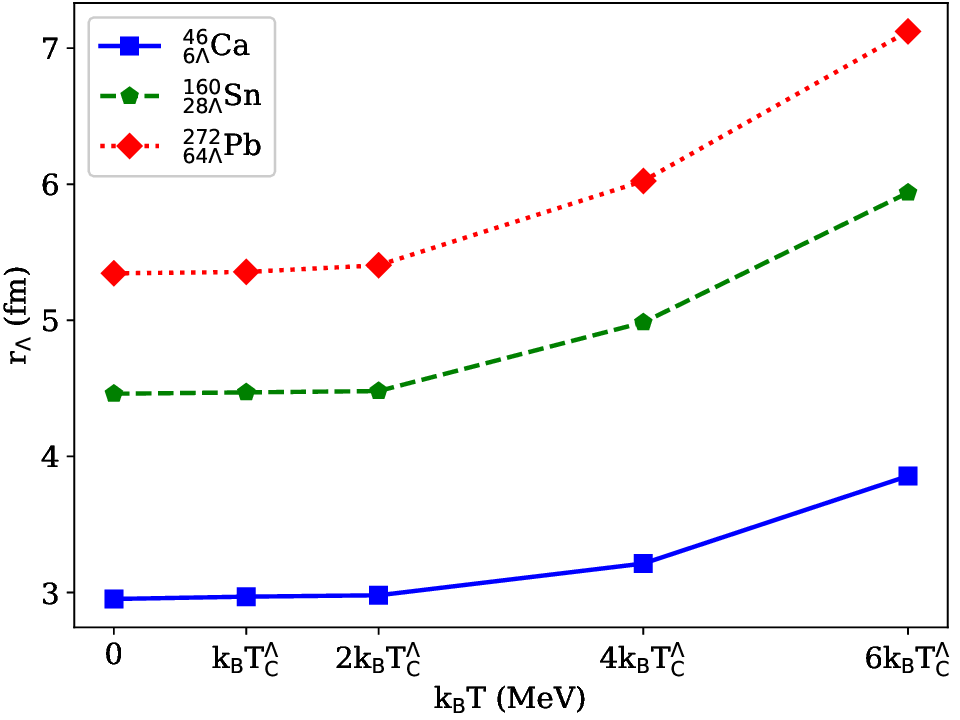}
\caption{Correlation between temperature and $\Lambda$ hyperon radii in hypernuclei $^{46}_{6\Lambda}$Ca, $^{160}_{28\Lambda}$Sn, and $^{272}_{64\Lambda}$Pb using DF-NSC97a+EmpC functional. Corresponding $k_BT_C^\Lambda$ values of these hypernuclei are listed in Table~\ref{tb:critical}.}
\label{fig:radii}
\end{figure}

The $\Lambda$ hyperon radius is also analyzed. Fig.~\ref{fig:radii} illustrates this in the hypernuclei $^{46}_{6\Lambda}$Ca, $^{160}_{28\Lambda}$Sn, and $^{272}_{64\Lambda}$Pb using the DF-NSC97a+EmpC functional.
Small effect of temperature are found. 
Across all examined hyper-nuclei, the modification in radii remains around 0.3\% as temperature changes from 0 to $T_C^\Lambda$. Similarly, this change remains relatively small as temperature extends from 0 to $2 T_C^\Lambda$ for these hyper-nuclei. For instance, in the case of $^{40-S}_{-S \Lambda}$Ca hyper-isotopes, the radius increase is roughly 0.6\%, while for $^{132-S}_{-S\Lambda}$Sn hyper-isotopes, it reaches around 3\%. In the scenario of $\Lambda$ hyperon-rich $^{208-S}_{-S\Lambda}$Pb, this increase amounts to approximately 4\%. 
In conclusion, we found only weak impact of temperature on the $\Lambda$ hyperon radius.
Beyond 2T$_C$, the sudden increase of the radii  is due to the creation of a neutron gas, which is an spurious effect of using a finite-size box~\cite{PhysRevC.94.054325}. 

\begin{figure}
\includegraphics[width=0.4681\textwidth]{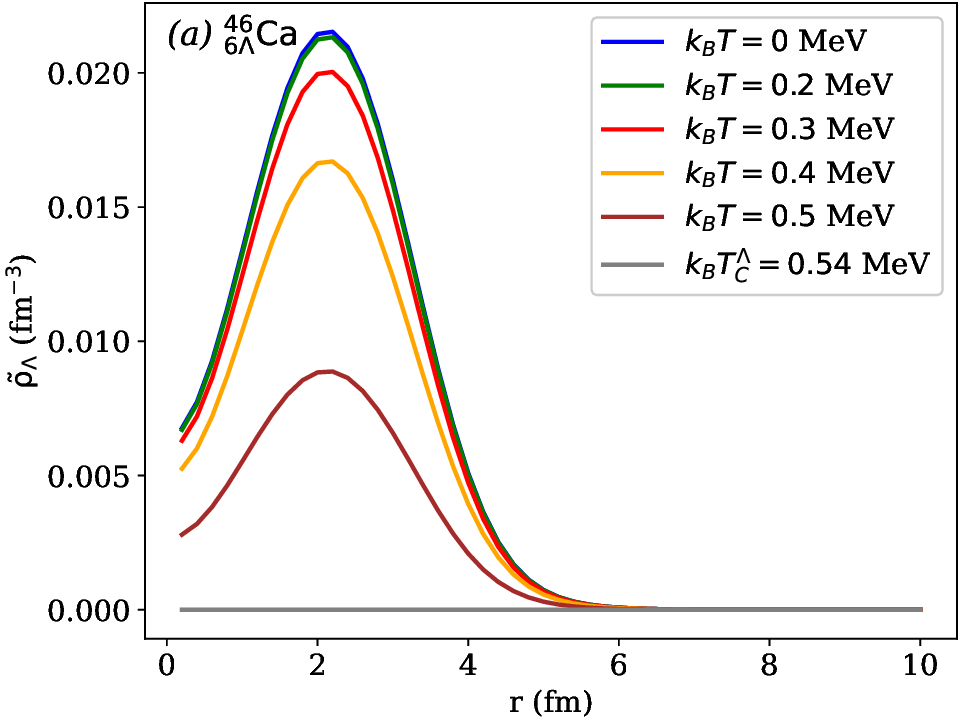}
\hspace{\fill}
\includegraphics[width=0.4681\textwidth]{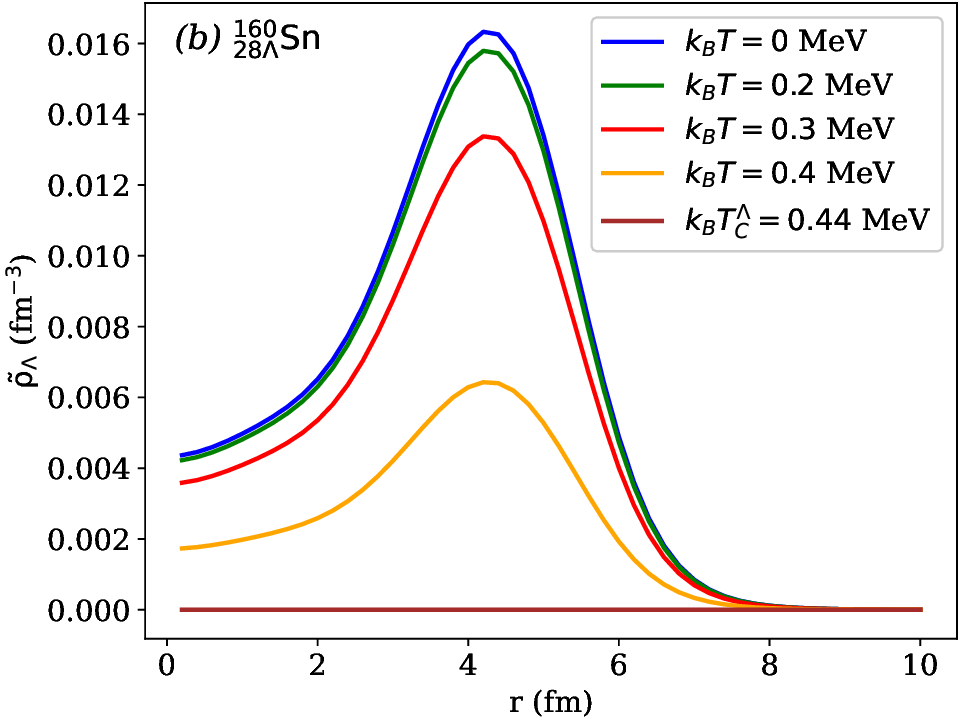}
\hspace{\fill}
\includegraphics[width=0.4681\textwidth]{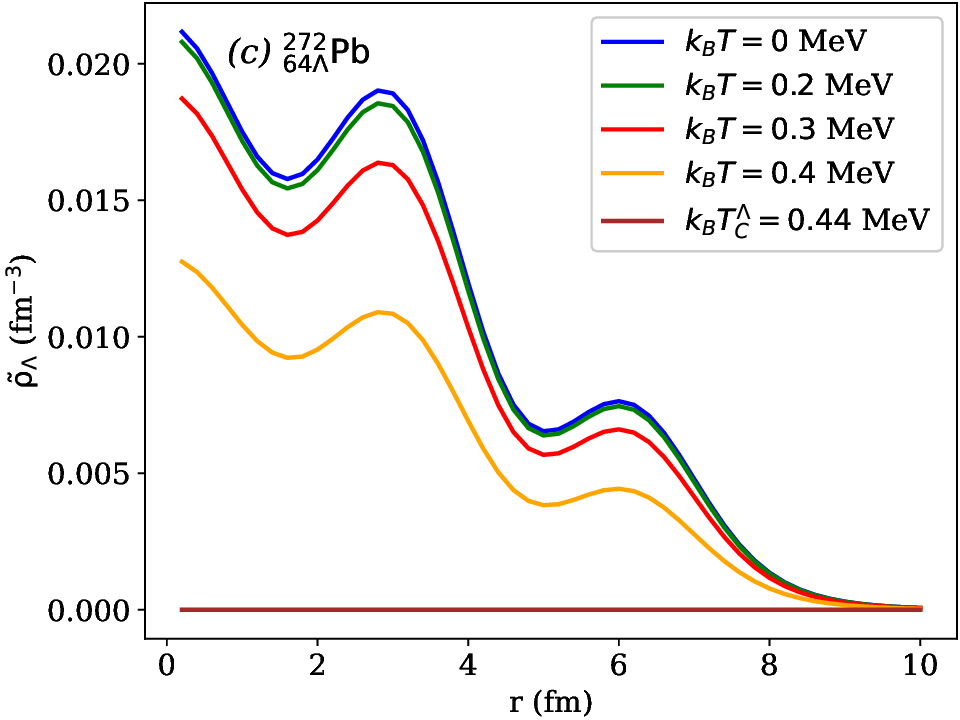}
\caption{Abnormal pairing density profiles for $^{46}_{6\Lambda}$Ca (a), $^{160}_{28\Lambda}$Sn (b), and $^{272}_{64\Lambda}$Pb (c) hypernuclei at finite temperatures using DF-NSC97a+EmpC functional.}
\label{fig:pairing_densities}
\end{figure}

While the impact of temperature on normal density profiles has been found to be negligible, the effect of temperature on the abnormal pairing density profiles is 
much stronger, as shown in Fig.~\ref{fig:pairing_densities}
for $^{46}_{6\Lambda}$Ca, $^{160}_{28\Lambda}$Sn, and $^{272}_{64\Lambda}$Pb hypernuclei.
Note that all the other hyper-nuclei that we have studied, see Fig.~\ref{fig:gap_97ae}, show similar results.
As expected, the reduction of the abnormal pairing density scales with the one of the pairing gaps shown in Fig.~\ref{fig:gap_97ae}.

\subsection{Pairing correlations around the hyperon-drip-line}

\begin{figure}                        
\includegraphics[width=0.4681\textwidth]{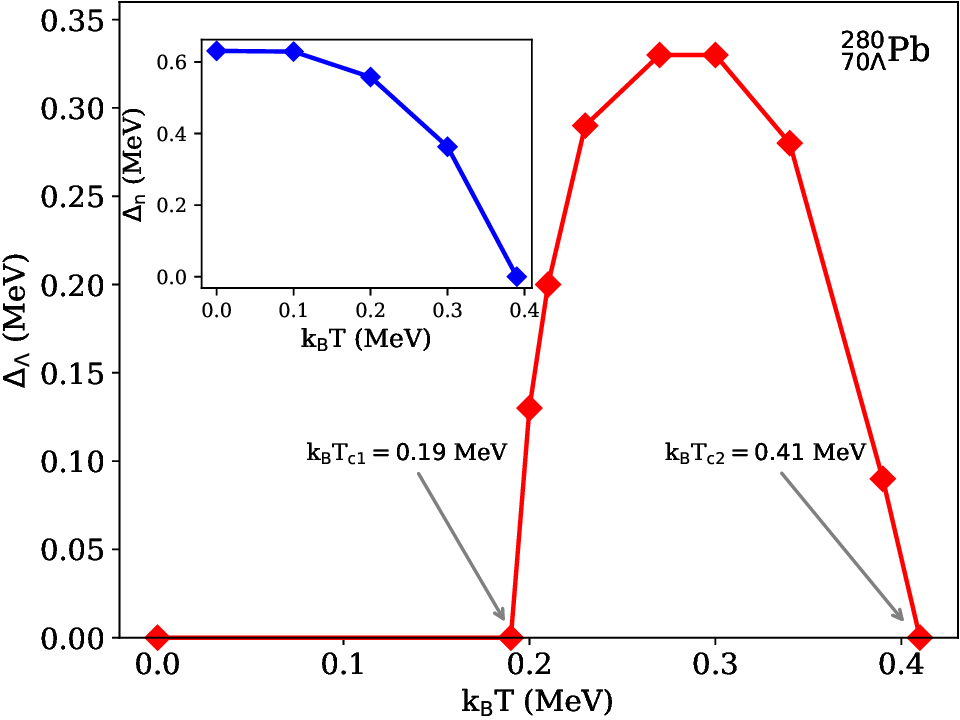}
\caption{Comparison of the evolution of $\Lambda$ and neutron pairing gaps with increasing temperature for the $^{280}_{70\Lambda}$Pb hypernucleus. The red line represents the $\Lambda$ pairing gap calculated using the DF-NSC97f+EmpC functional, while the blue line represents the neutron pairing gap using a pairing strength of $V_{n0} = -265$ MeV$\cdot$fm$^3$.}
\label{fig:reentrance}
\end{figure}

Among the $^{40-S}_{-S\Lambda}$Ca, $^{132-S}_{-S\Lambda}$Sn, and $^{208-S}_{-S\Lambda}$Pb hypernuclei that we examined in detail above, $^{60}_{20\Lambda}$Ca, $^{172}_{40\Lambda}$Sn, and $^{278}_{70\Lambda}$Pb lie the closest to the hyperon-drip-line at $T = 0$: their $\Lambda \Lambda$ separation energies is very close to 0. These drip-line systems can contribute to conditions where the pairing re-entrance effect (the pairing gap becomes non-zero above a given temperature) is predicted \cite{PhysRevC.86.065801, PhysRevC.96.024304}. However, as seen from Fig.~\ref{fig:gap_97ae}, even at finite temperatures, $\Delta_\Lambda = 0$ for these hypernuclei. This is not the case when considering open-shell neutron cases.  For this purpose, calculations were performed 
for $^{56,58,62,64}_{20\Lambda}$Ca, $^{168,170,174,176}_{40\Lambda}$Sn, and $^{274,276,280,282}_{70\Lambda}$Pb hypernuclei using all three functionals. Mixed-type pairing interaction was used in the neutron channel for these hypernuclei, with the pairing strength $V_{n0} = -265$ MeV$\cdot$fm$^3$, such that the pairing gap $\Delta_n\approx 0.6$~MeV in $^{280}_{70\Lambda}$Pb, see Fig.~\ref{fig:reentrance}. For the $\Lambda$ channel, we used the pairing strengthes given in Tab.~\ref{tb:strength}. Pairing re-entrance occurs in $^{280}_{70\Lambda}$Pb using DF-NSC97f+EmpC functional, with the two predicted critical temperatures $k_BT_{c1} = 0.19$ MeV and $k_BT_{c2} = 0.41$ MeV. It should be noted that at $k_BT_{c2} = 0.41$ MeV, using the DF-NSC97f+EmpC functional yields chemical potentials of -0.32 MeV for $\Lambda$ hyperons, -2.83 MeV for neutrons, and -7.21 MeV for protons in $^{280}_{70\Lambda}$Pb. However, for all other neutron open-shell hypernuclei we investigated, $\Delta_\Lambda$ remains constant at 0 across all temperatures. The reason for the pairing re-entrance phenomenon specific to $^{280}_{70\Lambda}$Pb, is that in this hypernucleus using DF-NSC97f+EmpC, the energy difference between the last occupied 3s level and the first unoccupied 1h level is smaller compared to other similar $^{274,276,278,282}_{70\Lambda}$Pb hypernuclei, as shown in Table~\ref{tb:reentrance}. As the temperature changes, the 1h level, which is for this hypernucleus empty at $T=0$, begins to fill, and pairing properties begin to emerge at temperatures greater than $k_B T = 0.19$ MeV.

\begin{table}
\tabcolsep=0.12cm
\def\arraystretch{1.4}
\caption{Energy difference between the highest occupied 3s level and the lowest unoccupied 1h level in $^{274,276,278,280,282}_{70\Lambda}$Pb hypernuclei at $T=0$.} \label{tb:reentrance}
\begin{tabular}{cccccc}
\hline \hline  \multicolumn{1}{c}{Functionals} & \multicolumn{5}{c}{ Energy difference $(\mathrm{MeV})$} \\
\cline { 2 - 6 } & $^{274}_{70\Lambda}$Pb & $^{276}_{70\Lambda}$Pb & $^{278}_{70\Lambda}$Pb & $^{280}_{70\Lambda}$Pb & $^{282}_{70\Lambda}$Pb\textbf{}\\
\hline DF-NSC89+EmpC & 1.87 & 1.79 & 1.81 & 1.80 & 1.76 \\
DF-NSC97a+EmpC  & 3.21 & 2.87 & 3.03 & 3.56 & 3.41 \\
DF-NSC97f+EmpC  & 1.68 & 1.68 & 1.59 & 1.53 & 1.62\\
\hline \hline
\end{tabular}
\end{table}

\section{Conclusions}
\label{sec:level4}

In conclusion, the interplay between pairing and finite temperature for multi-$\Lambda$ hypernuclei has been studied using the FT-HFB apparoach. We have considered three different functionals for the $N\Lambda$ channel, DF-NSC89, DF-NSC97a, and DF-NSC97f, all adjusted from microscopic Brueckner-Hartree-Fock predictions.
For the $\Lambda\Lambda$ channel, the empirical prescription EmpC has been used, which is adjusted to reproduce the experimental bond energy in $^6_{\Lambda\Lambda}$He.

The pairing gaps for individual hyper-nuclei show similar trends as the temperature approach the critical temperature $T_C^\Lambda$, which is found to be about $0.54$~MeV and to scale with respect to $\Delta_\Lambda(T=0)$ as suggested by the BCS relation. Among the studied functionals, DF-NSC97a+EmpC predicts the largest pairing gaps. 

We have also investigated the effect of temperature on the binding energy, on the single-particle spectra, on the normal density and on the $\Lambda$ hyperon radius. There is no effect of temperature for $T\lesssim T_C^\Lambda$. As expected, the abnormal pairing densities and the condensation energy are impacted by the temperature, and shrinks to zero for 
$T=T_C^\Lambda$. Calculations also explored hyperon-drip-line hypernuclei. A pairing re-entrance effect was observed in the neutron open-shell hyperon drip-line hypernucleus $^{280}_{70\Lambda}$Pb, in the $\Lambda $ channel.

In summary, this study clarifies the combined effects of temperature and pairing correlations in hyper-nuclei. 
Our main findings is to show that the usual BCS correlation between $T_C^\Lambda$ and $\Delta_\Lambda(T=0)$ remains valid for hyper-nuclei.

\begin{acknowledgments}
This work is supported by the Scientific and Technological Research Council of Turkey (T\"{U}B\.{I}TAK) under project number MFAG-118F098 and the Yildiz Technical University under project number FBI-2018-3325 and FBA-2021-4229.
This work is also supported by the LABEX Lyon Institute of Origins (ANR-10-LABX-0066) of the \textsl{Universit\'e de Lyon} and the IN2P3 masterproject MAC.
\end{acknowledgments}

\bibliography{paper}

\end{document}